\documentclass[11pt]{article}

\usepackage{amsmath}
\usepackage{amsthm}
\usepackage{amssymb}

\theoremstyle{plain}
\newtheorem{theorem}{Theorem}[section]
\newtheorem{lemma}[theorem]{Lemma}

\newtheorem{definition}[theorem]{Definition}

\newcommand{\abs}[1]{\left\lvert#1\right\rvert}
\DeclareMathOperator{\Sim}{sim}

\DeclareMathOperator{\Her}{Her}
\DeclareMathOperator{\HerQ}{Her_{\mathbb{Q}}}
\DeclareMathOperator{\tr}{tr}
\DeclareMathOperator{\diag}{diag}

\newcommand{\N}{\mathbb{N}}
\newcommand{\Z}{\mathbb{Z}}
\newcommand{\Q}{\mathbb{Q}}
\newcommand{\R}{\mathbb{R}}
\newcommand{\C}{\mathbb{C}}
\newcommand{\CQ}{\mathbb{C}_Q}
\newcommand{\X}{\Sigma^*}

\newcommand{\noi}{\noindent}

\title{\textbf{
Upper bound by Kolmogorov complexity for\\
the probability in computable POVM measurement
}}

\author{
Kohtaro Tadaki\\
\\
IMAI Quantum Computation and Information Project, ERATO,\\
Japan Science and Technology Corporation,\\
Daini Hongo White Bldg. 201,
5-28-3, Hongo, Bunkyo-ku, Tokyo 113-0033, Japan\\
TEL: +81-3-3818-3314 \hspace{2mm} FAX: +81-3-3818-3285\\
E-mail: tadaki@qci.jst.go.jp
}

\date{}

\begin{document}

\maketitle

\begin{quotation}
\noi\textbf{Abstract.}
We apply algorithmic information theory to quantum mechanics
in order to shed light on an algorithmic structure which
inheres in quantum mechanics.

There are two equivalent ways
to define the (classical) Kolmogorov complexity $K(s)$
of a given classical finite binary string $s$.
In the standard way,
$K(s)$ is defined as the length of the shortest input string
for the universal self-delimiting Turing machine to output $s$.
In the other way,
we first introduce the so-called universal probability $m$,
and then define $K(s)$ as $-\log_2 m(s)$
without using the concept of program-size.
We generalize the universal probability
to a matrix-valued function,
and identify this function with
a POVM (positive operator-valued measure).
On the basis of this identification,
we study a computable POVM measurement with countable measurement outcomes
performed upon a finite dimensional quantum system.
We show that,
up to a multiplicative constant,
$2^{-K(s)}$ is the upper bound
for the probability of each measurement outcome $s$
in such a POVM measurement.
In what follows,
the upper bound $2^{-K(s)}$ is shown to be optimal in a certain sense.
\end{quotation}

\vspace{1mm}

\begin{quotation}
\noi\textit{Key words\/}:
algorithmic information theory,
universal probability, POVM, computability,
quantum Kolmogorov complexity
\end{quotation}

\section{Introduction}

Algorithmic information theory is a theory of program-size complexity
which has precisely the formal properties of classical information theory.
In algorithmic information theory,
the \textit{program-size complexity} (or \textit{Kolmogorov complexity})
$K(s)$ of a finite binary string $s$
is defined as
the length of the shortest binary input
for the universal self-delimiting Turing machine to output $s$.
The concept of program-size complexity plays an important role in
characterizing the randomness of a finite or infinite binary string.
In this paper
we extend algorithmic information theory to quantum region
in order to throw light upon an algorithmic feature of quantum mechanics.
We show that Kolmogorov complexity gives the upper bound for
the probability of each measurement outcome
in a computable POVM measurement with countable outcomes
performed upon a finite dimensional quantum system.

\subsection{Main result}

In this paper,
we consider a quantum measurement
performed upon a \textit{finite dimensional} quantum system.
A \textit{positive operator-valued measure} (POVM) is
a collection $\{E(m)\}$ of positive semi-definite Hermitian matrices
which satisfies $\sum_{m}E(m)=I$
where $I$ is the identity matrix.
Each $E(m)$ is called a \textit{POVM element} of this POVM.
In general,
the statistics of outcomes in a quantum measurement are
described by a POVM $\{E(m)\}$.
The label $m$ refers to the measurement outcomes
that may occur in the experiment.
If the state of the quantum system is described
by a normalized vector $|\psi\rangle$
immediately before the measurement,
then the probability that result $m$ occurs is given by
$\langle\psi|E(m)|\psi\rangle$.
On the other hand,
if the ensemble of the states of the quantum system is described
by a density matrix $\rho$ immediately before the measurement,
then the probability that result $m$ occurs is given by
$\tr(\rho E(m))$.
A POVM measurement is a generalization of
a \textit{projective measurement}
which is described by an observable.
The number of outcomes in a POVM measurement can be more than
the dimension of the state space of the quantum system being measured,
whereas the number of outcomes in a projective measurement cannot.
In this paper,
we relate an argument $s$ of $K(s)$ to an outcome
which may occur in the quantum measurement
performed upon a finite dimensional quantum system.
Since $K(s)$ is defined for all finite binary strings $s$,
the countable outcomes have to be available
in the corresponding quantum measurement.
Thus we deal with a POVM measurement and not a projective measurement.
(See e.g. \cite{NC00,P00} for the details of POVM measurements.)

We say a POVM is \textit{computable}
if one can compute all its POVM elements to any desired degree of precision,
and a POVM measurement is said to be \textit{computable}
if it is described by a computable POVM.
Our main result is as follows:
Let $\{R(s)\}$ be a computable POVM on a finite dimensional quantum system
whose each element is labeled by a finite binary string.
Then there exists an integer $d$ such that,
for all density matrix $\rho$
and all finite binary string $s$,
\begin{equation}\label{K}
  K(s)-d \le -\log_2\tr(\rho R(s)),
\end{equation}
and also there exists a real number $c>0$ such that,
for all density matrix $\rho$
and all finite binary string $s$,
\begin{equation}\label{P}
   \tr(\rho R(s)) \le c\,P(s).
\end{equation}
Here $P(s)$ is the probability that
the (classical) universal self-delimiting Turing machine halts and outputs $s$
when it starts on the program tape filled with an infinite binary string
generated by infinitely repeated tosses of a fair coin.

The inequality \eqref{K} states that,
up to an additive constant,
$K(s)$ is the lower bound
for the $-\log_2$ of the probability of each measurement outcome $s$
in a computable POVM measurement with countable outcomes
performed upon a finite dimensional quantum system,
i.e., $2^{-K(s)}$ is the upper bound for the probability of
each outcome $s$ up to a multiplicative constant.
On the other hand,
the inequality \eqref{P} states that,
up to a multiplicative constant,
$P(s)$ is the upper bound
for the probability of each measurement outcome $s$
in the same measurement.
Note that
the inequalities \eqref{K} and \eqref{P} are equivalent to each other.

The computability of a POVM measurement is thought to be intrinsic
in the case where one performs the measurement
in order to extract a valuable information from a quantum system
because in such a case
one has to be able to compute to any desired degree of precision
all POVM elements of the POVM which describes the measurement.
Hence,
when one wants to extract a valuable information
from a finite dimensional quantum system through a POVM measurement
with countable outcomes,
one faces with the limitation given by the inequality \eqref{K}
(equivalently by \eqref{P}).

Especially,
the inequality \eqref{P} is interesting.
Since $P(s)$ is a probability
which results from infinitely repeated tosses of a fair coin,
$P(s)$ is just a classical probability.
In the case where $\rho$ is a pure state,
the inequality \eqref{P} states that
a purely quantum mechanical probability bounded from above by
a purely classical probability up to a multiplicative constant
when one performs a computable POVM measurement
with countable outcomes
upon a finite dimensional quantum system
in the pure state $\rho$.

The inequalitis \eqref{K} and \eqref{P} are obtained
through a generalization of the so-called \textit{universal probability}
to a matrix-valued function.
The Kolmogorov complexity $K(s)$ of a finite binary string $s$
is originally defined using the concept of program-size.
However,
there is another way to define $K(s)$ without referring to such a concept,
that is,
we first introduce a universal probability $m$,
and then define $K(s)$ as $-\log_2 m(s)$.
The universal probability is a function
from the set of finite binary strings to the open interval $(0,1)$.
In this paper
we generalize the universal probability to a matrix-valued function
while keeping the domain of definition the set of finite binary strings.
Then this generalized universal probability is identified with
an analogue of a POVM,
and is called a \textit{universal semi-POVM}.
The inequalities \eqref{K} and \eqref{P} naturally
follow from this identification.

\subsection{Related works}

Our aim is to generalize algorithmic information theory
in order to understand the algorithmic feature of quantum mechanics.
There are related works whose purpose is mainly
to define the information content of an individual pure quantum state,
i.e.,
to define the \textit{quantum Kolmogorov complexity} of the quantum state
\cite{V01,BVL01,G01},
while we will not make such an attempt in this paper.

As we mentioned above,
$K(s)$ can be defined as 
the $-\log_2$ of the universal probability
without using the concept of program-size.
\cite{G01} took this approach
in order to define
the information content
of a pure quantum state.
\cite{G01} first generalized the universal probability to
a matrix-valued function $\boldsymbol{\mu}$,
called \textit{quantum universal semi-density matrix}.
The $\boldsymbol{\mu}$ is a function
which maps any positive integer $N$ to
an $N\times N$ positive semi-definite Hermitian matrix $\boldsymbol{\mu}(N)$
with its trace less than or equal to one.
\cite{G01} proposed to regard $\boldsymbol{\mu}(N)$
as an analogue of a density matrix of a quantum system
called \textit{semi-density matrix}.
Then,
in order to measure
the information content of
a pure quantum state $|\psi\rangle\in\C^N$,
\cite{G01} introduced the \textit{quantum algorithmic entropies}
$\underline{H}(|\psi\rangle)$
and
$\overline{H}(|\psi\rangle)$
as
$-\log_2\langle\psi|\boldsymbol{\mu}(N)|\psi\rangle$
and
$-\langle\psi|(\log_2\boldsymbol{\mu}(N))|\psi\rangle$,
respectively.
In general,
the trace of a density matrix has to be equal to one.
If the trace of $\boldsymbol{\mu}(N)$ is equal to one,
then the quantity $\langle\psi|\boldsymbol{\mu}(N)|\psi\rangle$
in the definition of $\underline{H}(|\psi\rangle)$ has the meaning of
the probability that the outcome is `yes'
when one performs the projective measurement described by
the projector $|\psi\rangle\langle\psi|$
upon the quantum system in the mixed state $\boldsymbol{\mu}(N)$.
However,
the trace of $\boldsymbol{\mu}(N)$ is not
equal to one for all but finitely many $N$
because of its universality.
(This fact is implicitly mentioned in \cite{G01}.
For completeness,
we include a proof of this fact in Appendix \ref{pr_trace<1},
in addition to the definition of $\boldsymbol{\mu}$.)

In quantum mechanics,
what is represented by a matrix is
either a quantum state or a measurement operator.
In this paper
we generalize the universal probability to a matrix-valued function
in different way from \cite{G01},
and identify it with an analogue of a POVM.
We do not stick to defining the information content of a quantum state.
Instead,
we focus our thoughts on
applying algorithmic information theory to quantum mechanics
in order to shed light on an algorithmic structure of quantum mechanics.
In this line we have the above inequalities \eqref{K} and \eqref{P}.

In each of \cite{V01} and \cite{BVL01},
the quantum Kolmogorov complexity of a qubit string was defined
as a quantum generalization of the standard definition of
classical Kolmogorov complexity;
the length of the shortest input
for the universal decoding algorithm $U$
to output a finite binary string.
Both \cite{V01} and \cite{BVL01} adopt
the \textit{universal quantum Turing machine}
as a universal decoding algorithm $U$ to output a quantum state
in their definition.
However,
there is a difference between \cite{V01} and \cite{BVL01}
with respect to the object which is allowed as an input to $U$.
That is,
\cite{V01} can only allow a classical binary string as an input,
whereas \cite{BVL01} can allow any qubit string.
The works \cite{V01}, \cite{BVL01}, and \cite{G01} are
closely related to one another
as shown in each of these works.
In comparison with our work,
since our work is, in essence,
based on a generalization of the universal probability,
the work \cite{G01} is more related to our work
than the works \cite{V01} and \cite{BVL01}.
These two works may be related to our work
via the work \cite{G01}.

\subsection{Organization of the paper}

We begin in Section \ref{preliminaries}
with some basic definitions,
and review some results of algorithmic information theory.
In Section \ref{pr_main},
we prove the inequalities \eqref{K} and \eqref{P}
via the introduction of a universal semi-POVM
as a generalization of the universal probability.
In Section \ref{on_optimality},
we consider the optimality of our upper bound $2^{-K(s)}$ and $P(s)$
for the probability of each measurement outcome $s$.
Finally,
we study some other properties of a universal semi-POVM
in Section \ref{prop-usP}.

\section{Preliminaries}
\label{preliminaries}

\subsection{Notation}

We start with some notation about numbers and matrices
which will be used in this paper.

$\N \equiv \left\{0,1,2,3,\dotsc\right\}$
is the set of natural numbers,
and $\N^+$ is the set of positive integers.
$\Z$ is the set of integers.
$\Q$ is the set of rational numbers,
and $\Q^+$ is the set of positive rational numbers.
$\R$ is the set of real numbers, and $\C$ is the set of complex numbers.
$\CQ$ is the set of the complex numbers in the form of $a+ib$ with
$a,b\in\Q$.
We define $-\log_2 0$ as $\infty$.

We fix $N$ to be any one positive integer throughout this paper.
For each matrix $A$, $A^T$ is the transpose of $A$
and $A^\dagger$ is the adjoint of $A$.
For each $K\subset\C$,
$M_N(K)$ is the set of the $N\times N$ matrices whose elements are in $K$,
and $K^N$ is the set of column vectors consist $N$ complex numbers in $K$.
For each $x=(x_1,x_2,\dots,x_N)^T\in\C^N$,
$\|x\|$ is defined as
$(\abs{x_1}^2+\abs{x_2}^2+\dots+\abs{x_N}^2)^{1/2}$.
For each $A, B\in M_N(\C)$,
$[A,B]$ is defined as $AB-BA$.
For each $A\in M_N(\C)$,
$\|A\|$ is the \textit{operator norm} of $A$,
and $\tr A$ denotes the \textit{trace} of $A$.
The \textit{identity matrix} in $M_N(\C)$ is denoted by $I$.
$U(N)$ is the set of $N\times N$ unitary matrices.
$\Her(N)$ is the set of $N\times N$ Hermitian matrices.
For each $A, B\in\Her(N)$,
we write $A\leqslant B$ if $B-A$ is positive semi-definite,
and write $A < B$ if $B-A$ is positive definite.
Note that the relation $\leqslant$ on $\Her(N)$ is a partial order.
In this paper we will frequently use the property:
$\|A\|\le\varepsilon \Longleftrightarrow
-\varepsilon I \leqslant A \leqslant \varepsilon I$
for any $\varepsilon\ge 0$ and any $A\in\Her(N)$.
We say $\rho$ is a \textit{density matrix} if
$0\leqslant\rho\in\Her(N)$ and $\tr(\rho)=1$.
$\HerQ(N)$ is the set of $N\times N$ Hermitian matrices
whose elements are in $\CQ$.
$\diag(x_1,\dots,x_N)$ is the diagonal matrix
whose $(i,i)$-elements is $x_i$.

Let $S$ be any set,
and let $f,g\colon S\to\Her(N)$.
Then we write
$f(x)=g(x)+O(1)$
if there is a real number $c>0$ such that,
for all $x\in S$,
$\|f(x)-g(x)\|\le c$.
We also write
$f(x)\sim g(x)$
if there is a real number $c>0$ such that,
for all $x\in S$,
$c\,f(x)\leqslant g(x)$ and $c\,g(x)\leqslant f(x)$.

$\X \equiv
\left\{
  \lambda,0,1,00,01,10,11,000,001,010,\dotsc
\right\}$
is the set of finite binary strings
where $\lambda$ denotes the \textit{empty string},
and $\X$ is ordered as indicated.
We identify any string in $\X$ with a natural number in this order,
that is,
we consider $\varphi\colon \X\to\N$ such that $\varphi(s)=1s-1$
where the concatenation $1s$ of strings $1$ and $s$ is regarded
as a dyadic integer,
and then we identify $s$ with $\varphi(s)$.
For any $s \in \X$, $\abs{s}$ is the \textit{length} of $s$.
A subset $S$ of $\X$ is called a \textit{prefix-free set}
if no string in $S$ is a prefix of another string in $S$.

For each $F\colon \X\to M_N(\C)$,
we say $F$ is \textit{computable} if
there exists a total recursive function
$G\colon \X\times\N\to M_N(\CQ)$ such that,
for all $s\in \X$ and all $k\in\N$,
$\|F(s)-G(s,k)\| < 2^{-k}$.

\subsection{Algorithmic information theory}

In the following
we review some definitions and results of
algorithmic information theory \cite{C75,C87a}.
We assume that the reader is familiar with algorithmic information theory
in addition to computability theory.

A \textit{computer} is a partial recursive function
$C\colon \X\to \X$ whose domain of definition
is a prefix-free set.
For each computer $C$ and each $s \in \X$,
$K_C(s)$ is defined as
$\min
\left\{\,
  \abs{p}\,\big|\;p \in \X\>\&\>C(p)=s
\,\right\}$.
A computer $U$ is said to be \textit{optimal} if
for each computer $C$ there exists a constant $\Sim(C)$
with the following property:
if $C(p)$ is defined, then there is a $p'$ for which
$U(p')=C(p)$ and $\abs{p'}\le\abs{p}+\Sim(C)$.
It is then shown that there exists a computer which is optimal.
We choose any one optimal computer $U$
as the standard one
for use throughout the rest of this paper,
and we define
$K(s) \equiv K_U(s)$,
which is referred to as
the \textit{information content} of $s$,
the \textit{program-size complexity} of $s$, or
the \textit{Kolmogorov complexity} of $s$.
For each $s\in \X$,
$P(s)$ is defined by
$P(s)\equiv\sum_{U(p)=s}2^{-\abs{p}}$.
The class of computers is equal to the class of functions
which are computed by \textit{self-delimiting Turing machines}.
A self-delimiting Turing machine has a program tape and
a work tape.
The program tape is infinite to the right,
while the work tape is inifinite in both directions.
The machine starts with an input string on its program tape
and the work tape blank.
When the machine halts, the output string is put on the work tape.
(For the details of self-delimiting Turing machine, see \cite{C75}.)
A self-delimiting Turing machine is called \textit{universal} if
it computes an optimal computer.
Let $M_U$ be a universal self-delimiting Turing machine which computes $U$.
Then $P(s)$ is the probability that $M_U$ halts and outputs $s$
when $M_U$ starts on the program tape filled with an infinite binary string
generated by infinitely repeated tosses of a fair coin.

A universal probability is defined through the following two definitions.
\begin{definition}\label{lcsp}
  For any $r\colon \X\to[0,\infty)$,
  we say that $r$ is a lower-computable semi-measure if
  $r$ satisfies the following two conditions:
  \begin{enumerate}
    \item $\sum_{s\in \X}r(s)\le 1$.
    \item There exists a total recursive function $f\colon\N\times \X\to\Q$
      such that, for each $s\in \X$,
      $\lim_{n\to\infty} f(n,s)=r(s)$ and
      $\forall\,n\in\N\;\>f(n,s)\le f(n+1,s)$.
  \end{enumerate}
\end{definition}
\begin{definition}
  Let $m$ be a lower-computable semi-measure.
  We say that $m$ is a universal probability if
  for any lower-computable semi-measure $r$,
  there exists a real number $c>0$ such that,
  for all $s\in \X$, $c\,r(s)\le m(s)$.
\end{definition}

Then the following theorem holds.

\begin{theorem}\label{eup}
  Both $2^{-K(s)}$ and $P(s)$ are universal probabilities.
\end{theorem}

By Theorem \ref{eup}, we see that, for any universal probability $m$,
\begin{equation}\label{eq: K_m}
  K(s)=-\log_2 m(s)+O(1).
\end{equation}
Especially we have $K(s)=-\log_2 P(s)+O(1)$.
Any universal probability is not computable,
which corresponds to the uncomputability of $K(s)$.
Moreover we can show the following,
from which the uncomputability of a universal probability follows.

\begin{theorem}\label{upper-uncomputabe_up}
  Let $m$ be a universal probability,
  and let $f\colon \N\to\Q^+$ and
  $\tau\colon\N\to\X$.
  Suppose that both $f$ and $\tau$ are total recursive functions,
  and $m(\tau(n))\le f(n)$ for all $n\in\N$.
  Then $\inf_{s\in \X}f(n)>0$.
\end{theorem}

The information theoretic feature of algorithmic information theory
can be developed as follows.
We choose any one computable bijection $<s,t>$
from $(s,t)\in \X\times \X$ to $\X$.
Let $s, t\in \X$.
The \textit{joint information content} $K(s,t)$ of $s$ and $t$ is defined as
$K(s,t)\equiv K(<s,t>)$.
We then define
the \textit{relative information content} $K(s|t)$ of $s$ relative to $t$
by the equation
\begin{equation*}
  K(s|t)\equiv K(t,s)-K(t).
\end{equation*}
Finally we define the \textit{mutual information content}
$K(s:t)$ of $s$ and $t$
by the equation
\begin{equation*}
  K(s:t)\equiv K(t)-K(t|s) \equiv K(s)+K(t)-K(s,t).
\end{equation*}
Then, without referring to the concept of program-size,
\cite{C87a} proved the following relations using the fact that
$2^{-K(s)}$ is a universal probability.

\begin{theorem}\label{rel_AIT}\
  \begin{enumerate}
    \item $K(s,t)=K(t,s)+O(1)$.
    \item $K(s:t)=K(t:s)+O(1)$.
    \item $K(s:s)=K(s)+O(1)$.
    \item $\exists\,c\in\R\;\>\forall\,s,t\in \X\;\>c\le K(s|t)$.
    \item $\exists\,c\in\R\;\>\forall\,s,t\in \X\;\>c\le K(s:t)$.
    \item $K(s:t)=K(t:s)+O(1)$.
    \item $K(s:s)=K(s)+O(1)$.
    \item $K(s:\lambda)=O(1)$.
  \end{enumerate}
\end{theorem}

Thus algorithmic information theory has
the formal properties of classical information theory.

\section{Generalization of universal probability to POVM}
\label{pr_main}

In this section
we generalize a universal probability to a matrix-valued function.
Based on this generalization,
we prove our main result: Theorem \ref{computable-povm-mixed}.

\begin{definition}\label{def-povm}
  We say $R$ is a semi-POVM on $\X$ if
  $R$ is a mapping from $\X$ to $\Her(N)$ which satisfies
  $0\leqslant R(s)$ for all $s\in \X$ and
  $\sum_{s\in \X}R(s)\leqslant I$.
  We say $R$ is a POVM on $\X$ if $R$ is semi-POVM on $\X$
  and $\sum_{s\in \X}R(s) = I$.
\end{definition}

Let $Q$ be a POVM on $\X$.
The POVM measurement described by $Q$
is performed upon a finite dimensional quantum system,
and gives one of countable measurement outcomes,
which are represented by finite binary strings.

Given $R$: semi-POVM on $\X$,
it is easy to convert $R$ into a POVM on $\X$
by appending an appropriate positive semi-definite matrix to $R$.
Let $\Omega=\sum_{s\in\X}R(s)$,
and then we define $Q\colon\X\to\Her(N)$ by
$Q(\lambda)=I-\Omega$ and $Q(s')=R(s)$ for each $s\in\X$
where $s'$ is the successor of $s$.
Then $Q$ is a POVM on $\X$.
Thus a semi-POVM on $\X$ has a physical meaning
in the same way as a POVM on $\X$.

\begin{definition}\label{def-cpovm}
  We say $R$ is a lower-computable semi-POVM if
  $R$ is a semi-POVM on $\X$
  and there exists a total recursive function
  $f\colon\N\times \X\to\HerQ(N)$
  such that, for each $s\in \X$,
  $\lim_{n\to\infty} f(n,s) = R(s)$ and
  $\forall\,n\in\N\;\>f(n,s) \leqslant R(s)$.
\end{definition}

In the case where $N=1$,
Definition \ref{def-cpovm} exactly results in
the definition of a lower-computable semi-measure.
For the handiness,
we do not reqiure in the above definition that
the $f(n,s)$ conversing to $R(s)$ is
non-decreasing (i.e., $f(n,s)\leqslant f(n+1,s)$).
However,
we can equivalently assume that the $f(n,s)$ is non-decreasing
in the definition.
See Appendix \ref{pr_itiretuka} for its proof.

The following is a key theorem for our main result.

\begin{theorem}\label{main}
  If $R$ is a lower-computable semi-POVM,
  then the mapping $\X\ni s\longmapsto \frac{1}{N}\|R(s)\|$ is
  a lower-computable semi-measure.
\end{theorem}

\begin{proof}
  Let $r\colon \X\to[0,\infty)$ with $r(s)=\frac{1}{N}\|R(s)\|$.
  Note that $\|A\|\le \tr A$ for any positive semi-definite $A$.
  Thus,
  since $0\leqslant R(s)$ for all $s\in \X$
  and $\sum_{s\in \X}R(s)\leqslant I$,
  we see that
  $\sum_{s\in \X} r(s) \le
  \frac{1}{N} \tr \sum_{s\in \X} R(s) \le
  \frac{1}{N} \tr I = 1$. Thus the condition (i) in Definition \ref{lcsp}
  holds for $r$.

  Next we show that
  the condition (ii) in Definition \ref{lcsp} holds for $r$.
  Since $R$ is a lower-computable semi-POVM,
  there exists a total recursive function
  $f\colon\N\times \X\to\HerQ(N)$
  such that for each $s\in \X$,
  $\lim_{n\to\infty} f(n,s)$ $=R(s)$ and
  $\forall\,n\in\N\;\>
  f(n,s)\leqslant R(s)$.
  From the definition of the operator norm,
  $\|f(n,s)\|$ is the supremum of
  $\langle\psi|f(n,s)|\psi\rangle/\langle\psi|\psi\rangle$
  such that $|\psi\rangle\neq 0$ and $|\psi\rangle\in\C^N$.
  Since $\Q$ is dense in $\R$,
  it is easy to see that
  $\|f(n,s)\|$ is equal to the supremum of
  $\langle\psi|f(n,s)|\psi\rangle/\langle\psi|\psi\rangle$
  such that $|\psi\rangle\neq 0$ and
  each component of $|\psi\rangle$ is a complex number
  in the form of $a+ib$ with $a,b\in\Z$.
  Thus, given $n\in\N$ and $s\in \X$,
  one can generate a sequence of rational numbers $p_1,p_2,\dotsc$
  such that
  $p_1\le p_2\le \dots \le\|f(n,s)\|$ and
  $\lim_{m\to\infty} p_m = \|f(n,s)\|$.
  On the other hand,
  using the property $A\leqslant B\Longrightarrow\|A\|\le \|B\|$,
  we have $\|f(n,s)\|\le\|R(s)\|$
  and $\lim_{n\to\infty} \|f(n,s)\|$ $=\|R(s)\|$.
  Hence, given $s\in \X$,
  one can generate a sequence of rational numbers $x_1,x_2,\dotsc$
  such that
  $x_1\le x_2\le \dots \le \|R(s)\|$ and
  $\lim_{n\to\infty} x_n = \|R(s)\|$.
  Therefore the condition (ii) in Definition \ref{lcsp} holds for $r$.
  Hence $r$ is a lower-computable semi-measure.
\end{proof}

\begin{definition}\label{def-suP}
  Let $M$ be a lower-computable semi-POVM.
  We say that $M$ is a universal semi-POVM if
  for each lower-computable semi-POVM $R$,
  there exists a real number $c>0$ such that
  for all $s\in \X$, $c\,R(s)\leqslant M(s)$.
\end{definition}

In the case where $N=1$,
Definition \ref{def-suP} exactly results in
the definition of a universal probability.
The use of the partial order $\leqslant$ for the purpose of generalizing
lower-computable semi-measure and universal probability to
matrix-valued functions is suggested in \cite{G01}.
Note that if $M$ is a universal semi-POVM
then, for all $s\in \X$, $M(s)$ is positive definite.

A universal semi-POVM may have a simple form as the following
theorem says.

\begin{theorem}\label{exist-suP}
  If $m$ is a universal probability,
  then the mapping $\X\ni s\longmapsto m(s)I$ is
  a universal semi-POVM.
\end{theorem}

\begin{proof}
  Let $M\colon \X\to\Her(N)$ with $M(s)=m(s)I$.
  Since $m$ is a lower-computable semi-measure,
  it is obvious that $M$ is a lower-computable semi-POVM.
  Suppose that $R$ is a lower-computable semi-POVM.
  By Theorem \ref{main},
  the mapping $\X\ni s\longmapsto \frac{1}{N}\|R(s)\|$ is
  a lower-computable semi-measure.
  Thus, since $m$ is a universal probability,
  there is $c>0$ such that, for all $s\in \X$,
  $c\frac{1}{N}\|R(s)\|\le m(s)$.
  Therefore
  we have $\frac{c}{N}R(s) \leqslant m(s)I$ for all $s\in \X$.
  Hence $M$ is a universal semi-POVM.
\end{proof}

For this universal semi-POVM $m(s)I$,
we have $[m(s)I,m(t)I]=0$ for all $s$ and $t\in \X$.
However the following theorem guarantees an existence of
a `non-trivial' universal semi-POVM.

\begin{theorem}
  There exists a universal semi-POVM $M$ such that
  $[M(s),M(t)]\neq 0$ for any distinct $s$ and $t\in \X$.
\end{theorem}

\begin{proof}
  We choose any one universal probability $m$,
  and choose any one pair of $G$ and $H \in \HerQ(N)$ such that
  $0<G,H\leqslant I$ and $[G,H]\neq 0$.
  We define $M\colon \X\to\Her(N)$ by
  \begin{equation*}
    M(s)=
    \frac{m(s)}{2}
    \left(2^{-\varphi(s)}G+H\right).
  \end{equation*}
  Then we see that, for any distinct $s$ and $t\in \X$,
  \begin{equation*}
    [M(s),M(t)]=\frac{1}{4}
    m(s)m(t)\left(2^{-\varphi(s)}-2^{-\varphi(t)}\right)[G,H]\neq 0.
  \end{equation*}
  Since $m$ is a lower-computable semi-measure,
  $M$ is shown to be a lower-computable semi-POVM.
  It follows from $0<H$ that there is $c>0$ such that $cI\leqslant H$.
  Thus $\frac{c}{2}m(s)I\leqslant M(s)$.
  Since $m(s)I$ is a universal semi-POVM,
  $M$ is also a universal semi-POVM.
\end{proof}

The following theorem is more general form of our main result.

\begin{theorem}\label{upper bound}
  Let $m$ be a universal probability,
  and let $R$ be a lower-computable semi-POVM.
  Then the following (i) and (ii) hold:
  \begin{enumerate}
    \item There exists $c>0$ such that,
      for any normalized $|\psi\rangle\in\C^N$
      and any $s\in \X$,
      \begin{equation*}
        \langle\psi|R(s)|\psi\rangle \le c\,m(s).
      \end{equation*}
    \item There exists $c>0$ such that,
      for any density matrix $\rho\in\Her(N)$
      and any $s\in \X$,
      \begin{equation*}
        \tr(\rho R(s)) \le c\,m(s).
      \end{equation*}
  \end{enumerate}
\end{theorem}

\begin{proof}
  It follows from Theorem \ref{exist-suP} that (i) holds.
  Using (i) and the spectral decomposition of $\rho$,
  we have (ii). 
\end{proof}

In order to make more clear
the physical implication of Theorem \ref{upper bound},
we restrict our attention to a POVM on $\X$ which is computable.
Informally, a POVM on $\X$ is computable if and only if
one can compute all its POVM elements to any desired degree of precision.
Thus the computability of a POVM is thought to be inherent
in the case where one wants to perform a well-controlled quantum measurement
described by the POVM.
Using the following lemma,
we have our main result about a computable POVM.

\begin{lemma}\label{computable-semi-computable}
  Let $R$ be a semi-POVM on $\X$.
  If $R$ is computable then $R$ is a lower-computable semi-POVM.
\end{lemma}

\begin{proof}
  Since $R$ is computable,
  there exists a total recursive function
  $G\colon \X\times\N\to M_N(\CQ)$ such that,
  for all $s\in \X$ and all $k\in\N$,
  $\|R(s)-G(s,k)\| < 2^{-k}$.
  We define $H\colon \X\times\N\to M_N(\CQ)$ by
  $H(s,k) =
  \frac{1}{2}\left\{G(s,k)+G(s,k)^\dagger\right\}$.
  Then $H$ is a total recursive function and,
  for every $s\in \X$ and every $k\in\N$,
  $H(s,k)\in\HerQ(N)$ and $\left\|R(s)-H(s,k)\right\| < 2^{-k}$.
  Thus we have
  $H(s,k)-2^{-k}I\leqslant R(s)$ and
  $\lim_{k\to\infty} H(s,k)-2^{-k}I=R(s)$.
  Hence the result follows.
\end{proof}

\begin{theorem}[Main result]\label{computable-povm-mixed}
  Let $R$ be a computable POVM on $\X$.
  Then the following hold:
  \begin{enumerate}
    \item There exists $d\in\N$ such that,
      for any density matrix $\rho\in\Her(N)$
      and any $s\in \X$,
      \begin{equation}
        K(s)-d \le -\log_2\tr(\rho R(s)).
      \end{equation}
    \item There exists $c>0$ such that,
      for any density matrix $\rho\in\Her(N)$
      and any $s\in \X$,
      \begin{equation}\label{q-le-c}
        \tr(\rho R(s)) \le c\,P(s).
      \end{equation}
  \end{enumerate}
\end{theorem}

\begin{proof}
  Theorem \ref{computable-povm-mixed} immediately follows
  from Theorem \ref{eup}, (ii) in Theorem \ref{upper bound},
  and Lemma \ref{computable-semi-computable}.
\end{proof}

\section{Optimality of universal semi-POVM}
\label{on_optimality}

In this section
we consider an optimality of a universal semi-POVM.
By Theorem \ref{exist-suP}
we have the following theorem.

\begin{theorem}\label{M_sim_mI}
  Let $M$ be a universal semi-POVM.
  and let $m$ be a universal probability.
  Then $M(s)\sim m(s)I$.
\end{theorem}

The following theorem immediately follows
from Theorem \ref{M_sim_mI}.
This theorem is the most general form which represents
the optimality of a universal-semi POVM
from the view point of
the probability of each measurement outcome.

\begin{theorem}\label{optimality}
  Let $m$ be a universal probability,
  and let $M$ be a universal semi-POVM.
  Then
  there exist $c_1>0$ and $c_2>0$ such that,
  for any density matrix $\rho\in\Her(N)$ and any $s\in \X$,
  \begin{equation*}
    c_1 m(s) \le \tr(\rho M(s)) \le c_2 m(s). 
  \end{equation*}
\end{theorem}

By Theorem \ref{eup} and Theorem \ref{optimality},
we have Theorem \ref{optimality-k-p}.

\begin{theorem}\label{optimality-k-p}
  Let $M$ be a universal semi-POVM.
  Then, for any density matrix $\rho\in\Her(N)$ and any $s\in \X$,
  \begin{align*}
    K(s) &= -\log_2 \tr(\rho M(s))+ O(1), \\
    P(s) &\sim \tr(\rho M(s)).
  \end{align*}
\end{theorem}

Thus, if we can perform the POVM measurement described
by a universal semi-POVM,
then we can achieve the upper bound $P(s)$
(or $2^{-K(s)}$)
in Theorem \ref{computable-povm-mixed}
up to a multiplicative constant.
However
any universal semi-POVM is not computable
(see Subsection \ref{rel_to_up}).
Moreover we can show that
there is no computable semi-POVM on $\X$
which can achieve the upper bound $P(s)$ (or $2^{-K(s)}$)
up to a multiplicative constant.
Instead, by the definition of universal semi-POVM,
we have the following theorem,
which states that
we can approximate any universal semi-POVM
by a recursive sequence of semi-POVMs on $\X$ from below.

\begin{theorem}\label{approximation}
  For any universal semi-POVM $M$,
  there exists a sequence $F_0,F_1,F_2,\dotsc$ of semi-POVMs on $\X$
  such that
  \begin{enumerate}
    \item $F_n(s)\in\HerQ(N)$ and
      $0 < F_n(s) \leqslant F_{n+1}(s) \leqslant M(s)$
      for all $(n,s)\in\N\times \X$,
    \item the sequence $F_0,F_1,F_2,\dotsc$ of functions
      uniformly converges to $M$, and
    \item the mapping $\N\times\X\ni (n,s)\longmapsto F_n(s)$ is
      a total recursive function.
  \end{enumerate}
\end{theorem}

\begin{proof}
  Since $M$ is a universal semi-POVM,
  by Theorem \ref{itiretuka} in Appendix \ref{pr_itiretuka},
  there exists a total recursive function
  $g\colon\N\times \X\to\HerQ(N)$
  such that, for each $s\in \X$,
  $\lim_{n\to\infty} g(n,s) = M(s)$ and
  $\forall\,n\in\N\;\>g(n,s) \leqslant g(n+1,s) \leqslant M(s)$.
  Note that $0<M(s)$ for any $s\in\X$.
  Thus, there exists a total recursive function
  $\tau\colon\N\times\X\to\N$ such that,
  for each $s$ and $n$,
  $\tau(n,s)<\tau(n+1,s)$ and $0<g(\tau(n,s),s)$.
  We define the sequence $F_0,F_1,F_2,\dotsc$ of semi-POVMs on $\X$
  by $F_n(s)=g(\tau(n,s),s)$.
  It is then obvious that
  (i) and (iii) in Theorem \ref{approximation} hold for this sequence.
  For any $\varepsilon>0$,
  there is $s_0\in\X$ such that $\sum_{s>s_0} M(s)<\varepsilon I$,
  so we see that $\|F_n(s)-M(s)\|<\varepsilon$
  for all $n\in\N$ and all $s>s_0$.
  On the other hand,
  it is easy to see that there is $n_0\in\N$ such that,
  for all $n>n_0$ and all $s\le s_0$,
  $\|F_n(s)-M(s)\|<\varepsilon$.
  Thus (ii) in Theorem \ref{approximation} holds for
  the sequence $F_0,F_1,F_2,\dotsc$ of functions.
\end{proof}

For the recursive sequence $F_0,F_1,F_2,\dotsc$ of semi-POVMs on $\X$
given in Theorem \ref{approximation},
$F_n$ is a computable semi-POVM on $\X$ for each $n\in\N$.
However,
since any universal semi-POVM is not a POVM on $\X$
(see Subsection \ref{rel_to_up})
and $F_n(s) \leqslant M(s)$,
$F_n$ is not a POVM on $\X$ for each $n$.
Instead,
we can also consider the recursive sequence $G_1, G_2, G_3, \dotsc$ of POVMs
defined as follows:
Each POVM element of $G_n$ is labeled by a finite binary string
less than or equal to $\varphi(n)$.
For any $s<\varphi(n)$, $G_n(s)$ is defined as $F_n(s)$,
and $G_n(\varphi(n))$ is defined as $I-\sum_{s<\varphi(n)}F_n(s)$.
Then,
since $F_n(s)\in\HerQ(N)$,
any given $n\in\N^+$,
one can calculate all POVM elements of $G_n$.
Note that the POVM measurement described by $G_n$
gives one of $n+1$ measurement outcomes,
each of which is represented
by a finite binary string less than or equal to $\varphi(n)$.
By Theorem \ref{approximation},
we have the following:
\begin{itemize}
  \item Any given $\varepsilon>0$,
    for all sufficiently large $n\in\N^+$,
    if $s<\varphi(n)$ and $\rho$ is a density matrix
    then $0\le\tr(\rho M(s))-\tr(\rho G_n(s))<\varepsilon$.
\end{itemize}
Thus,
in the sense that the above statement holds,
the recursive sequence $G_1, G_2, G_3, \dots, G_n, \dotsc$ of POVMs
converges to the universal semi-POVM $M$ from below as $n\to\infty$.

\section{Other properties of universal semi-POVM}
\label{prop-usP}

In this section
we study the properties of universal semi-POVM further.

\subsection{Matrix-valued algorithmic information theory}

Let $M$ be any one universal semi-POVM.
The equation \eqref{eq: K_m}
suggests defining a matrix-valued Kolmogorov complexity $\mathcal{K}(s)$
of $s\in\X$ by
\begin{equation}\label{KM}
  \mathcal{K}(s)\equiv -\log_2 M(s).
\end{equation}
For this definition of $\mathcal{K}$,
it follows from Theorem \ref{M_sim_mI} that
\begin{equation}\label{mvK}
  \mathcal{K}(s)=K(s)I+O(1).
\end{equation}
Further we can define
$\mathcal{K}(s,t)$, $\mathcal{K}(s|t)$, and $\mathcal{K}(s:t)$
in the same manner as the definitions of $K(s,t)$, $K(s|t)$, and $K(s:t)$,
respectively.
Then using \eqref{mvK}
we see that all the relations in Theorem \ref{rel_AIT} hold
for these $\mathcal{K}$'s in place of the $K$'s.

Note that $K(s)$ is originally defined using the concept of program-size.
Since $\mathcal{K}(s)$ is related to $K(s)$ through the equation \eqref{mvK},
$\mathcal{K}(s)$ have the meaning of program-size in some weak sense.
It is interesting
if we can find a more concrete definition of $\mathcal{K}(s)$
using something like the concept of program-size
instead of the equation \eqref{KM}.
However,
this is still open.

In order to measure the information content of
a quantum state $|\psi\rangle\in\C^N$,
\cite{G01} introduced the quantum algorithmic entropies
$\underline{H}(|\psi\rangle)$
and
$\overline{H}(|\psi\rangle)$ of $|\psi\rangle$
as
$-\log_2\langle\psi|\boldsymbol{\mu}(N)|\psi\rangle$
and
$-\langle\psi|(\log_2\boldsymbol{\mu}(N))|\psi\rangle$,
respectively,
using his quantum universal semi-density matrix $\boldsymbol{\mu}$
(see Appendix \ref{pr_trace<1} for its definition).
In this behalf note that,
for our universal semi-POVM $M$,
the following holds for any normalized $|\psi\rangle\in\C^N$:
\begin{equation*}
  K(s)=-\log_2\langle\psi|M(s)|\psi\rangle + O(1)
  =-\langle\psi|(\log_2 M(s))|\psi\rangle + O(1).
\end{equation*}
Thus
$-\log_2\langle\psi|M(s)|\psi\rangle$ and
$-\langle\psi|(\log_2 M(s))|\psi\rangle$
are independent of $|\psi\rangle$
up to an additive constant.
So it would seem difficult to measure the information
content of a quantum state $|\psi\rangle$ using these quantities
in the similar manner to \cite{G01},
although such an attempt is not the purpose of this paper.

\subsection{Relation to universal probability}
\label{rel_to_up}

We say $x\in\C^N$ is \textit{computable} if
each component of $x$ is in the form of $a+ib$
where $a$ and $b$ are computable real numbers.
Theorem \ref{th:n-1-suP} describes a property of
a universal semi-POVM as a universal probability.

\begin{theorem}\label{th:n-1-suP}
  Let $M$ be a universal semi-POVM,
  and let $x\in\C^N$ be computable with $\|x\|=1$.
  Then the mapping $\X\ni s\longmapsto x^\dagger M(s)x$ is
  a universal probability.
\end{theorem}

\begin{proof}
  Since $x$ is computable,
  $x^\dagger M(s)x$ is shown to be a lower-computable semi-measure.
  Let $m$ be a universal probability.
  Then, by Theorem \ref{M_sim_mI},
  we see that $x^\dagger M(s)x\sim m(s)$.
  Thus the result follows.
\end{proof}

Let $M$ be a universal semi-POVM.
Then, by Theorem \ref{th:n-1-suP},
each diagonal element $M_{ii}(s)$ of $M(s)$
is a universal probability as a function of $s$,
and $\frac{1}{N}\tr(M(s))$ is also a universal probability
as a function of $s$.
Since any universal probability is not computable,
any one diagonal element $M_{ii}(s)$ is not computable.
Hence any universal semi-POVM is not computable.
If $M$ is a POVM on $\X$,
then,
since $M$ is a lower-computable semi-POVM,
we can show that $M$ is computable.
Thus any universal semi-POVM is not a POVM on $\X$.

\subsection{Computable unitary invariance}

We say $A\in M_N(\C)$ is \textit{computable} if
each element of $A$ is in the form of $a+ib$
where $a$ and $b$ are computable real numbers.
The following theorem states an invariance of a POVM measurement
described by a universal semi-POVM
under computable unitary transformation
on the quantum state being measured.

\begin{theorem}\label{unitary}
  Let $M$ be a universal semi-POVM,
  and let $U\in U(N)$ be computable.
  Then the mapping $\X\ni s\longmapsto U^\dagger M(s)U$ is
  a universal semi-POVM.
\end{theorem}

\begin{proof}
  We note the property that
  $A \leqslant B \Longrightarrow X^\dagger A X \leqslant X^\dagger B X$
  for any $A, B\in\Her(N)$ and any $X\in M_N(\C)$.
  Since $U$ is computable,
  $U^\dagger M(s)U$ is shown to be a lower-computable semi-POVM.
  Let $m$ be a universal probability.
  Then, by Theorem \ref{M_sim_mI},
  we see that $U^\dagger M(s)U\sim m(s)I \sim M(s)$.
  Thus the result follows.
\end{proof}

Let $U\in U(N)$ be a computable,
and let $\mathcal{M}$ be a POVM measurement
described by a universal semi-POVM.
Suppose that, any given state $\rho$,
we first evolve $\rho$ by the unitary transformation $U$,
and then perform the measurement $\mathcal{M}$ for the transformed state
(i.e., $U\rho U^\dagger$).
Then, by Theorem \ref{unitary},
the whole POVM measurement for $\rho$ is shown to be still described
by a universal semi-POVM.


\section*{Acknowledgments}

The author is grateful to Hiroshi Imai and Keiji Matsumoto
for their support.


\appendix
\section{Quantum universal semi-density matrix}
\label{pr_trace<1}

We reproduce the definition of quantum universal semi-density matrix
from \cite{G01} as follows.

\begin{definition}
  Let $\sigma\colon\N^+\to\bigcup_{N\ge 1}\Her(N)$.
  We say $\sigma$ is a lower semicomputable semi-density matrix if
  $\sigma$ satisfies the following conditions:
  \begin{enumerate}
    \item For each $N\in\N^+$,
      $0\leqslant \sigma(N)\in\Her(N)$ and $\tr(\sigma(N))\le 1$.
    \item There exists a total recursive function
  $f\colon\N^+\times \N\to\bigcup_{N\ge 1}\HerQ(N)$
  such that, for each $N\in\N^+$,
  $\lim_{k\to\infty} f(N,k) = \sigma(N)$ and
  $\forall\,k\in\N\;\>f(N,k)\in\HerQ(N)\;\&\;f(N,k) \leqslant f(N,k+1)$.
  \end{enumerate}
\end{definition}

\begin{definition}
  Let $\boldsymbol{\mu}$ be a lower semicomputable semi-density matrix.
  We say $\boldsymbol{\mu}$ is a quantum universal semi-density matrix
  if for any lower semicomputable semi-density matrix $\sigma$,
  there exists a real number $c>0$ such that, for all $N\in\N^+$,
  $c\,\sigma(N) \leqslant \boldsymbol{\mu}(N)$.
\end{definition}

\begin{theorem}\label{trace<1}
  If $\boldsymbol{\mu}$ is a quantum universal semi-density matrix,
  then $\tr(\boldsymbol{\mu}(N))<1$ for all but finitely many $N\in\N^+$.
\end{theorem}

\begin{proof}
  Since $\boldsymbol{\mu}$ is a lower semicomputable semi-density matrix,
  there exists a total recursive function
  $f$ on $\N^+\times \N$
  such that, for each $N\in\N^+$,
  $\lim_{k\to\infty} f(N,k) = \boldsymbol{\mu}(N)$ and
  $\forall\,k\in\N\;\>f(N,k)\in\HerQ(N)\;\&\;
  f(N,k)\leqslant\boldsymbol{\mu}(N)$.
  Let $\boldsymbol{\mu}_{ii}(N)$ be
  the $(i,i)$-element of $\boldsymbol{\mu}(N)$,
  and let $f_{ii}(N,k)$ be
  the $(i,i)$-element of $f(N,k)$.
  Then, since $\tr(\boldsymbol{\mu}(N))\le 1$,
  we see that
  $\boldsymbol{\mu}_{ii}(N)\le 1-\sum_{j\neq i}f_{jj}(N,k)$.
  Especially,
  for any $N$ with $\tr(\boldsymbol{\mu}(N))=1$,
  we have
  $\boldsymbol{\mu}_{ii}(N)=\lim_{k\to\infty}1-\sum_{j\neq i}f_{jj}(N,k)$.
  On the other hand,
  it follows from $\sum_{i=1}^{N}\boldsymbol{\mu}_{ii}(N)\le 1$
  that $\min\{\boldsymbol{\mu}_{ii}(N)\mid 1\le i\le N\}\le 1/N$.
  Therefore,
  any given $\varepsilon>0$,
  for each sufficiently large $N$,
  there is $i$ such that
  $1\le i\le N$ and $\boldsymbol{\mu}_{ii}(N)<\varepsilon$.

  Now, contrary to Theorem \ref{trace<1},
  let us assume that,
  for infinitely many $N\in\N^+$,
  $\tr(\boldsymbol{\mu}(N))=1$.
  Then, given $\varepsilon\in\Q^+$,
  by checking for each $(N,i,k)$ in an exhaustive order
  whether $1-\sum_{j\neq i}f_{jj}(N,k)<\varepsilon$ holds or not,
  one can find $(N,i)$ such that
  $\boldsymbol{\mu}_{ii}(N)<\varepsilon$.
  Let $m$ be any one universal probability,
  and we define the lower semicomputable semi-density matrix $\sigma$
  by $\sigma(N)=\diag(m(\varphi^{-1}(1)),\dots,m(\varphi^{-1}(N)))$.
  Then, since $\boldsymbol{\mu}$ is a quantum universal semi-density matrix,
  for this $\sigma$,
  there is $c_{\sigma}>0$ such that
  if $1\le i\le N$
  then $c_{\sigma}m(\varphi^{-1}(i))\le \boldsymbol{\mu}_{ii}(N)$.
  It follows that
  there exists a total recursive function $\tau\colon\N\to\X$ such that,
  for any $n\in\N$, $m(\tau(n))\le 2^{-n}$.
  This contradicts Theorem \ref{upper-uncomputabe_up}.
  Thus we have Theorem \ref{trace<1}.
\end{proof}

\section{On the definition of lower-computable semi-POVM}
\label{pr_itiretuka}

The following theorem guarantees that
one can equivalently assume that $f(n,s)$ converging to $R(s)$
is non-decreasing in Definition \ref{def-cpovm}.

\begin{theorem}\label{itiretuka}
  $R$ is a lower-computable semi-POVM if and only if
  $R$ is a semi-POVM on $\X$
  and there exists a total recursive function
  $f\colon\N\times \X\to\HerQ(N)$
  such that for each $s\in \X$,
  $\lim_{n\to\infty} f(n,s) = R(s)$ and
  $\forall\,n\in\N\;\>f(n,s) \leqslant f(n+1,s)$.
\end{theorem}

For the proof of Theorem \ref{itiretuka} we need the following lemma,
which is an elementary result of linear algebra.

\begin{lemma}\label{principal_minors}
  For any $A\in\Her(N)$,
  $0\leqslant A$ if and only if
  all principal minors of $A$ are non-negative.
\end{lemma}

By Lemma \ref{principal_minors},
given $A$ and $B$ in $\HerQ(N)$,
one can effectively check whether $A \leqslant B$ holds or not.

\begin{proof}[Proof of Theorem \ref{itiretuka}]
  Assume that $R$ is a semi-POVM on $\X$
  and there exists a total recursive function
  $f\colon\N\times \X\to\HerQ(N)$
  such that for each $s\in \X$,
  $\lim_{n\to\infty} f(n,s) = R(s)$ and
  $\forall\,n\in\N\;\>f(n,s) \leqslant R(s)$.
  Let $g\colon\N\times \X\to\HerQ(N)$ be a total recursive function
  such that $g(n,s)=f(n,s)-2^{-n}I$.
  Then, for each $s\in \X$,
  $\lim_{n\to\infty} g(n,s) = R(s)$ and
  $\forall\,n\in\N\;\>g(n,s) < R(s)$.
  Thus,
  for each $s$ and $n$, there is a positive real number $c$ such that
  $cI \leqslant R(s)-g(n,s)$,
  and then, for this $c$, there is an $m\in\N$ such that $m>n$ and
  $R(s)-g(m,s) \leqslant cI$. So we have $g(n,s) \leqslant g(m,s)$.
  Thus,
  given $s$ and $n$,
  by checking $g(n,s) \leqslant g(k,s)$ for each $k>n$ in increasing order,
  one can finally find an $m$ with $g(n,s) \leqslant g(m,s)$.
  Therefore there exists a total recursive function
  $\tau\colon\N\times\X\to\N$ such that,
  for each $s$ and $n$,
  $\tau(n,s)<\tau(n+1,s)$ and $g(\tau(n,s),s) \leqslant g(\tau(n+1,s),s)$.
  We define a total recursive function $h\colon\N\times \X\to\HerQ(N)$
  by $h(n,s)=g(\tau(n,s),s)$.
  Then, for each $s\in \X$,
  $\lim_{n\to\infty} h(n,s) = R(s)$ and
  $\forall\,n\in\N\;\>h(n,s) \leqslant h(n+1,s)$.
  Hence, $R$ is a lower-computable semi-POVM.

  The other implication is obvious.
  Thus the theorem is obtained.
\end{proof}


\begin{thebibliography}{14}
\bibitem{BV97}
  Bernstein E. and Vazirani U.,\;
  Quantum complexity theory,
  \textit{SIAM J. Comput.},
  \textbf{26} (1997), 1411--1473.
\bibitem{BVL01}
  Berthiaume A., van Dam W., and Laplante S.,\;
  Quantum Kolmogorov complexity,
  \textit{J. Compute. System Sci.},
  \textbf{63} (2001), 201--221.
\bibitem{B96}
  Bhatia R.,\;
  \textit{Matrix Analysis\/},
  Springer, New York, 1996.
\bibitem{CHKW01}
  Calude C. S., Hertling P. H., Khoussainov B., and Wang Y.,\;
  Recursively enumerable reals and Chaitin $\Omega$ numbers,
  \textit{Theoret. Comput. Sci.},
  \textbf{255} (2001), pp.125--149.
\bibitem{C02}
  Calude C. S.,\;
  \textit{Information and Randomness: An Algorithmic Perspective\/},
  2nd Edition, Revised and Extended,
  Springer, Berlin, 2002.
\bibitem{C75}Chaitin G. J.,\;
  A theory of program size formally identical to
  information theory,
  \textit{J. Assoc. Comput. Mach.},
  \textbf{22} (1975), pp.329--340.
\bibitem{C87a}Chaitin G. J.,\;
  Incompleteness theorems for random reals,
  \textit{Adv. in Appl. Math.}, \textbf{8} (1987), pp.119--146.
\bibitem{G01}G\'acs P.,\;
  Quantum algorithmic entropy,
  \textit{J. Phys. A: Math. Gen.}, \textbf{34} (2001), pp.6859--6880.
\bibitem{HJ85}Horn R. A. and Johnson C. R.,\;
  \textit{Matrix Analysis\/},
  Cambridge University Press, Cambridge, 1985.
\bibitem{KS01}Ku\v{c}era A. and Slaman T. A.,\;
  Randomness and recursive enumerability,
  \textit{SIAM J. Comput.}, \textbf{31} (2001), pp.199--211.
\bibitem{NC00}Nielsen M. A. and Chuang I. L.,\;
  \textit{Quantum Computation and Quantum Information\/},
  Cambridge University Press, Cambridge, 2000.
\bibitem{P00}Preskill J.,\;
  \textit{Quantum Computation\/}, 2000.
  Course notes available at URL:
  \verb#http://www.theory.caltech.edu/people/preskill/ph229/#.
\bibitem{V01}Vit\'anyi P. M. B.,\;
  Quantum Kolmogorov complexity based on classical descriptions,
  \textit{IEEE Trans. Inform. Theory}, \textbf{47} (2001), pp.2464--2479.
\bibitem{LV97}
  Li M. and Vit\'anyi P. M. B.,\;
  \textit{An Introduction to Kolmogorov Complexity and Its Applications\/},
  Second Edition,
  Springer, New York, 1997.
\end{thebibliography}
\end{document}